\documentclass[aps,prl,twocolumn]{revtex4}

\usepackage{graphicx}
\usepackage{dcolumn}
\usepackage{bm}
\newcommand{\etal}{\emph{et al.}}

\begin{document}

\title{Reply to ``Comment on `Insulating Behavior of $\lambda$-DNA on the Micron Scale' "}

\author{Y. Zhang$^1$, R. H. Austin$^1$, E. C. Cox$^2$ and N. P. Ong$^1$}      
\affiliation{
Department of Physics$^1$, and Department of Molecular Biology$^2$, Princeton University, Princeton, New Jersey 08544, U.S.A.}
\begin{abstract}
In our experiment~\cite{Z}, we found that the resistance of vacuum-dried $\lambda$-DNA exceeds $10^{14}\; \Omega$ 
at 295 K.  Bechhoefer and Sen~\cite{BS} have raised a number of objections to our conclusion.  We provide counter arguments
to support our original conclusion.
\end{abstract}
\maketitle
\begin{figure}[h]   
\includegraphics[width=6cm]{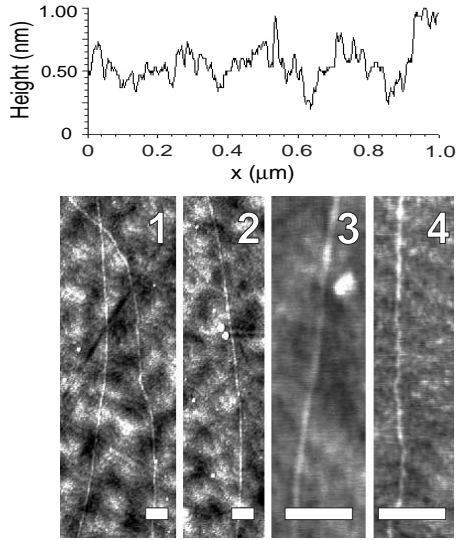}
\caption {\label{profile}  (a) Height profile of quartz substrate used~\cite{Z} (vertical scale expanded by $\sim$400).  (b) AFM images of DNA molecules on 
quartz taken with Molecular Imaging model PicoSPM II in tapping mode.  The probe is a silicon cantilever operating at a resonance frequency of 
$\sim$290 kHz (tip radius 10-20 nm).  Height scale: 1 nm from the darkest to the brightest. Scale bar: 200 nm.}
\end{figure}

Our experiment showed that the resistance of vacuum-dried $\lambda$-DNA exceeds $10^{14}\; \Omega$ at 295 K (resistivity $\rho_{DNA} >10^6\ 
\Omega$cm)~\cite{Z}.  We also found that salt residues can produce a spurious conductivity if not removed.  Bechhoefer and Sen~\cite{BS} comment that 
`events' that interrupt the conductivity invalidate our finding.  Surface roughness is mentioned as a problem.  However, AFM scans show that the height 
variance is $\sim 0.2-0.3$ nm over areas of 1 $\times$ 1 $\mu\mathrm{m}^2$ in our substrates (Fig. \ref{profile}a).  A larger variance ($\sim$ 2 nm) is 
observed on the Au surface.  Pitting on either Au or quartz surfaces has not been observed.  Hence our substrates are markedly smoother than 
assumed~\cite{BS}.  Figure~\ref{profile}b is a topographic image of some of the $\lambda$-DNA strands used in the experiment.  The majority of 
molecules laid down (Panels 1-3) are free of the kinks and sharp bends assumed by Ref.~\cite{BS}.  In rare cases, we do observe strands exhibiting 
sharp bends with radii of curvature 20-40 nm (Panel 4).  However, the evidence is that the chemical bonds defining the double-strand structure can 
withstand considerable tension.  

It is further envisaged~\cite{BS} that the Mg$^{2+}$ ions cause severe stretching.  Under gradual vacuum evaporation, the molecules should settle 
gradually into local potential minima.  Once they adhere, we do not see how the large forces implied in Ref.~\cite{BS} can arise.  Rivetti {\em et al.} have 
studied at length the binding of DNA to Mg$^{2+}$-modified mica surfaces~\cite{Bustamante}.  They find that strong surface charge may reduce the 
persistence length, but do not observe local stretching of the DNA contour length.

Yet another suggestion is that the reduced height creates insulating regions.  While our experiment was not designed to address this last effect, we point 
out that, to produce our $10^{14}\; \Omega$ value, the `events' envisaged must present barriers exceeding 80 mV (3 $k_BT$) to block conduction 
completely.  Otherwise, we should have observed a nonlinear current-voltage curve with our large bias.  

Height determination of DNA is not trivial.  The apparent height of DNA measured by AFM in our experiment $d_{obs} =$ 0.5 -1 nm, smaller than the 
expected 2 nm, but consistent with that found by other groups.  Muller and Engel~\cite{Muller} found that $d_{obs}$ in buffer solution depends on pH, 
electrolytic concentration and force applied.  For DNA in high vacuum, Uchihashi \etal~\cite{Uchihashi} show that DNA on a substrate maintains the B-form 
(with helical turns resolved), while $d_{obs}\sim$ 0.6 nm.  Finally, residual salts are the chief source of spurious high-conductivity readings~\cite{Z}, yet 
steps to remove them are rarely mentioned.  For these reasons, we feel that the link between stretching and conductivity is far from 
proven.

Several groups~\cite{Storm,Gomez,Heim} have inferred values of $\rho_{DNA} (10^5 - 10^6\; \Omega$cm) consistent with ours.  A previous observation 
of high microwave conductivity ($\rho_{DNA}\sim 1\;\Omega$cm) has now been traced to absorption by water molecules~\cite{Armitage}.

\end{document}